\documentclass[twocolumn]{aastex701}
\usepackage{booktabs}
\usepackage{array}
\usepackage{makecell}
\newcommand{\piso}{p_{\text{iso}}}
\newcommand{\ngas}{n_{\text{g}}}
\newcommand{\ncr}{n_{\text{CR}}}
\newcommand{\ecr}{\epsilon_{\text{CR}}}
\newcommand{\etacr}{\eta_{\text{CR}}}
\newcommand{\Emax}{\bar{E}_{\text{max}}}
\newcommand{\MA}{M_{\text{A}}}
\renewcommand{\arraystretch}{1.2} 
\usepackage{ragged2e}
\usepackage[hang,flushmargin]{footmisc}
\usepackage{amsmath}

\def \jpp {JPP}

\begin{document}
\title{Hybrid Simulations of Supersonic Shear Flows:  II) Cosmic Ray Viscosity}
\date{\today}

\author[0009-0003-0137-3116]{Naixin Liang}
\affiliation{Department of Physics, University of California, Santa Barbara, CA 93106, USA}
\email{naixin@ucsb.edu}

\author[0000-0003-0939-8775]{Damiano Caprioli}
\affiliation{Department of Astronomy \& Astrophysics, University of Chicago, Chicago, IL 60637, USA}
\email{luca.orusa@princedon.edu} 
\affiliation{Enrico Fermi Institute, The University of Chicago, Chicago, IL 60637, USA}

\begin{abstract}
In this second paper in a series dedicated to characterizing shear layers via 2D hybrid (kinetic ions -- fluid electrons) simulations, we study the dynamical role of nonthermal particles (cosmic rays, CRs), either spontaneously generated or pre-existing. 
We initialize Kolmogorov-type sinusoidal velocity shear flows unstable to the Kelvin--Helmholtz instability, which evolve nonlinearly into turbulence. 
Particles with large gyroradii act as long-range messengers that promote momentum exchange between layers, hence introducing a form of CR viscosity.
Even when not energetically dominant, increasing the CR energy density generally enhances momentum transfer, provided that their gyroradii are smaller than the shear lengthscale.
We consider flows ranging from subsonic to supersonic and assess the rate of shear dissipation, the partition of the initial kinetic energy among heating, thermal ion acceleration, CR reacceleration, and magnetic-field amplification, and the maximum energy attained by accelerated particles.
\end{abstract}

\section{Introduction}\label{sec:intro}

This series of papers uses kinetic plasma simulations to model the dissipation of collisionless, nonrelativistic, shear flows and the partitioning of their free kinetic energy.
In \cite{liang+26} (henceforth Paper I), we considered both subsonic and supersonic shear flows, studying the dissipation rate and the production of nonthermal particles as functions of the shear Mach number.
We found that supersonic flows generally lead to the formation of shocklets that produce energetic particles, which contribute significantly to the disruption of the shear.

Shear flows are very common in astrophysical environments such as in jet boundaries \citep[e.g.,][]{rieger+19}, accretion disks \citep[e.g.,][]{armitage11,belyaev+12}, gas motions in the turbulent interstellar medium (ISM) and intracluster medium \citep[e.g.,][]{zuhone+16,simionescu+19}.
We refer to Paper I for an extended discussion of possible applications.

In this paper, we consider the role that pre-existing nonthermal particles (or cosmic rays, CRs) have in inducing an effective kinetic viscosity in astrophysical collisionless plasmas.
The basic form of viscosity hinges on the momentum exchange mediated by the collisions between individual particles in their microscopic motions \citep{braginskii65}. 
However, atomic or molecular viscosity is usually negligible on astronomical scales, giving way to plasma instabilities and turbulence to dominate momentum transport \citep[e.g.,][]{balbus+98,liu+2006,kunz+11,schekochihin+22}.
CRs have large gyroradii and diffusion lengths, which allow them to couple distant plasma regions, effectively inducing a form of viscous momentum transport, as discussed in the pioneering paper by  \citet{earl+88}. 
They can also gain energy at the expense of the velocity gradients and the ensuing back-reaction on the thermal plasma produces an effective viscous stress \citep[e.g.][]{rieger+04,Webb_2018,Webb_2019}.
This physics can in principle also be embedded in fluid and magneto-hydrodynamical (MHD) approaches via a viscous stress tensor calibrated on the actual plasma collisionality.
However, in fluid approaches viscosity can only dissipate kinetic energy directly into heat, while in reality energy may also be channeled into freshly-accelerated particles, CR reacceleration, and magnetic fluctuations.
The mean CR energy generally increases because of second-order Fermi processes and, in turn, CRs can transfer energy and momentum to the other species.
\citet{earl+88} used a modified Parker transport equation in which the CR scattering rates and the viscous momentum diffusion coefficient were prescribed by hand, thereby gaining  some insights into the dynamical role of CRs, but the problem of the partitioning of the free kinetic energy of a shear inherently requires a kinetic approach that can provide the micro-physical momentum exchange rate due to strongly nonlinear wave--particle interactions.

We use hybrid particle-in-cell (PIC) simulations with kinetic ions and fluid electrons to model ab initio the partitioning of the shear free kinetic energy into turbulent motions, nonthermal particles, magnetic fields, and heat.
In Paper I we focused on the effects of the shear Mach number, spanning across subsonic and supersonic regimes; now, we consider the role that energetic particles with large gyroradii (CR ``seeds") have in inducing an effective kinetic viscosity in astrophysical collisionless plasmas.
Since in Paper I we showed that energetic particles can be spontaneously produced by the shear dissipation, the simulations presented here can also describe the long-term evolution of a system where shear flows are continuously triggered.

We initialize a sinusoidal velocity profile for the shear, known as a ``Kolmogorov flow" (also referred to as \textit{K-flow} in ergodic theory), immersed in a population of CRs as a test case to study turbulence and viscosity \citep[e.g.][]{meshalkin+61,thess92}.
The stability of Kolmogorov flows is primarily determined by the Reynolds number, with the Kelvin--Helmholtz instability setting in above a critical value \citep[e.g.,][]{chandra61, frank+96, miura97, oparina+04}.
In ideal magnetohydrodynamics (MHD), a sufficiently strong streamwise magnetic field can completely suppress the instability if the Alfv\'en speed exceeds the velocity shear across the layer \citep[e.g.,][]{hunt66, sommeria+82, gonzalez+94, chow+23, fraser+25}. 
In non-ideal regimes, processes such as magnetic resistivity or pressure anisotropy can trigger secondary microinstabilities, such as the firehose or gyrothermal instabilities, which further complicate the long-term evolution of the flow \citep[e.g.,][]{schekochihin+10, decamillis+15}.
Here, we do not focus on the mathematical properties of Kolmogorov flows, but use them as a convenient continuum shear that can be modeled with standard periodic boundary conditions, as opposed to the shearing boundary conditions used, e.g., by \citet{kunz+14b, sandoval+24}.

The paper is structured as follows: \S\ref{sec:setup} describes the simulation setup; \S\ref{sec:CRs} presents the results with different pre-existing CR number densities and energies; \S\ref{sec:sigmas} explores different Mach numbers of the shear flow; we  conclude in  \S\ref{sec:conclusion}.

\section{Simulation Setup}\label{sec:setup}
As in Paper I, we perform hybrid simulations of shear flows with {\tt dHybridR} \citep{haggerty+19a}, where ions are evolved as kinetic macro-particles under the relativistic Lorentz force and electrons are treated as a massless, charge-neutralizing fluid. 
The system is two-dimensional (2D in the $x$--$y$ plane) and 3D in both momentum and electromagnetic field components.

All physical quantities are normalized to their initial values.
The reference mass density is $\rho_0 = m_i n_0$, where $m_i$ denotes the ion (proton) mass and $n_0$ the ion number density.
Magnetic fields are normalized to $B_0$, velocities to the Alfv\'en speed $v_A = B_0 / \sqrt{4\pi \rho_{i,0}}$, and spatial scales to the ion inertial length $d_i = c / \omega_p$, with $c = 100v_A$ representing the speed of light and $\omega_p = \sqrt{4\pi n_0 e^2 / m_i}$ the ion plasma frequency.
Time is measured in units of the inverse ion cyclotron frequency, $\omega_c^{-1} = m_i c / (e B_0)$.
With these normalizations, the gyroradius of an ion with thermal speed $v_{\rm th,i} = v_A$ equals $d_i$, which also means that sonic and Alfv\'enic Mach numbers are equal to each other. 
Electrons are taken adiabatic ($P_e \propto \rho^{5/3}$) and thermal equilibrium with the ions \citep[see, e.g.,][for an extended discussion of this choice for supersonic flows]{caprioli+18}.

The simulation box is periodic in $x$ and $y$, with side lengths $L_x=L_y=L=200d_i$; 
we use two cells per $d_i$ and 100 particles per cell to ensure sufficient phase-space sampling. 
The timestep is fixed in $\delta t=2.5\times10^{-3}\omega_{c}^{-1}$, small enough that the fastest ions move less than one cell per time step. 
All simulations are evolved for several hundred ion cyclotron times ($\omega_{c}^{-1}$) to capture the decay of the shear and the ensuing nonthermal dynamics. Convergence with respect to spatial and temporal resolution and particle number has been verified.

While in Paper I we used a double shear layer following the benchmark of \citet{henri+13}, here we consider a Kolmogorov flow periodic in $y$ with velocity $\textbf{U}= U_{x}(y)\textbf{e}_{x}$, where:
\begin{equation}\label{eq:Ux}
U_x(y)\equiv U_0 \sin\left(\frac{2\pi y}{L}\right),
\end{equation}
This choice removes the parameter that defines the layer thickness (fixed to  $3d_i$ in Paper I) and is more representative of continuous shear layers.
Moreover, if the CR gyroradius $r_{\rm g}\gg d_i$, this setup allows us to span cases where $r_{\rm g}$ is either smaller or larger than the shear scale $L$, which correspond to different coupling regimes  between CRs and thermal plasma.

CR seeds are initialized with density $n_{\rm CR}$ and an isotropic momentum $p_{\rm iso}\equiv p/(m_iv_A)$ in their rest frame; 
then, they are boosted with the local sinusoidal bulk flow.
The initial magnetic field $\mathbf{B}_0=B_0\mathbf{e}_x$ is uniform, so CRs start scattering and thus coupling with the thermal plasma only when magnetic fluctuations are self-generated by the evolution of the shear flow.

Simulation runs are grouped according to the parameters varied, as in Table \ref{tab:param}. 
The $\mathcal{N}$, $\mathcal{P}$, and $\mathcal{E}$ series explore different CR number densities, momentum distributions, and energy densities, respectively, to assess how CR-induced viscosity modifies energy and momentum transfer in the shearing plasma.
The $\mathcal{L}$ run uses a larger simulation box. 
The $\mathcal{S}$ series (Runs $\mathcal{S}1 -\mathcal{S}10$) varies the initial Alfv\'enic Mach number to test how the shear amplitude affects the shear-reduction timescales. 
Depending on the initial CR content and shear strength, the total energy can be initially dominated by either CR or thermal pressure, as we discuss more in \ref{subsec:efrac}.
This parameter space is chosen not to represent the most common astrophysical situations possible (see Paper I), but rather to cover different regimes in terms of flow Mach numbers (from subsonic to supersonic) and the energetic predominance of either CRs or thermal plasma. 

Though the backreaction of CRs on the shear can be modeled also via MHD-PIC simulations with fluid background \citep[e.g.,][]{bai+15, liu+26}, only kinetic approaches can distinguish between heating and acceleration, besides capturing the kinetic microinstabilities that may provide effective viscosity \citep[][]{schekochihin+10, rosin+11}.

Although the parameter space is limited by the finite range of values treatable in a kinetic approach (which precludes, for instance, treating the CR/gas densities $\sim 10^{-9}$ typical of the ISM), we consider CR energy densities both above and below equipartition with the thermal plasma and CR gyroradii both larger and smaller than the flow gradient scale-height. 
Future work will expand this general analysis to values more tailored to specific astrophysical applications.

\begin{table}[htbp]
\centering
\begin{tabular}{|c|c|c|c|c|}
\hline
\parbox{1.0cm}{\centering Run} & \parbox{1.0cm}{\centering $M_{\rm A}$} & \parbox{1.5cm}{\centering $n_{\rm CR}$ [\%]} & \parbox{1.5cm}{\centering $p_{\rm iso} [m_iv_A]$} & \parbox{0.8cm}{\centering $L [d_i]$} \\
\hline
$\mathcal{B}$   & 20 & 1    & 200  & 200 \\
\hline
$\mathcal{N}1$  & 20 & 10   & 200  & 200 \\
$\mathcal{N}2$  & 20 & 5    & 200  & 200 \\
$\mathcal{N}3$  & 20 & 0.5  & 200  & 200 \\
$\mathcal{N}4$  & 20 & 0.1  & 200  & 200 \\
$\mathcal{N}5$  & 20 & 0    & -    & 200 \\
\hline
$\mathcal{P}1$  & 20 & 1    & 2000 & 200 \\
$\mathcal{P}2$  & 20 & 1    & 1000 & 200 \\
$\mathcal{P}3$  & 20 & 1    & 80   & 200 \\
$\mathcal{P}4$  & 20 & 1    & 50   & 200 \\
\hline
$\mathcal{E}1$  & 20 & 10   & 50   & 200 \\
$\mathcal{E}2$  & 20 & 5    & 50   & 200 \\
$\mathcal{E}3$  & 20 & 0.1  & 50   & 200 \\
\hline
$\mathcal{L}$   & 20 & 10   & 50   & 2000 \\
\hline
$\mathcal{S}1$  & 40 & 1 & 200 & 200 \\
$\mathcal{S}2$  & 30 & 1 & 200 & 200 \\
$\mathcal{S}3$  & 15 & 1 & 200 & 200 \\
$\mathcal{S}4$  & 12 & 1 & 200 & 200 \\
$\mathcal{S}5$  & 10 & 1 & 200 & 200 \\
$\mathcal{S}6$  & 8  & 1 & 200 & 200 \\
$\mathcal{S}7$  & 5  & 1 & 200 & 200 \\
$\mathcal{S}8$  & 2  & 1 & 200 & 200 \\
$\mathcal{S}9$  & 1  & 1 & 200 & 200 \\
$\mathcal{S}10$ & 0.5& 1 & 200 & 200 \\
\hline
\end{tabular}%

\caption{Simulation parameters. From left to right: maximum Alfv\'enic Mach number ($M_{\rm A}$, also set equal to the sonic Mach number), CR number density, isotropic CR momentum, and box size in units of $d_i$. All CRs drift with the same velocity shear as the gas ions. All runs use a constant timestep of $\delta t = 2.5\times10^{-3} \omega_{c}^{-1}$.}
\label{tab:param}
\end{table}

\section{The Contribution of CRs}\label{sec:CRs}
\subsection{Shear Dissipation Timescales}
\label{subsec:tau}
\begin{figure*}
\centering
\includegraphics[width=2.1\columnwidth]{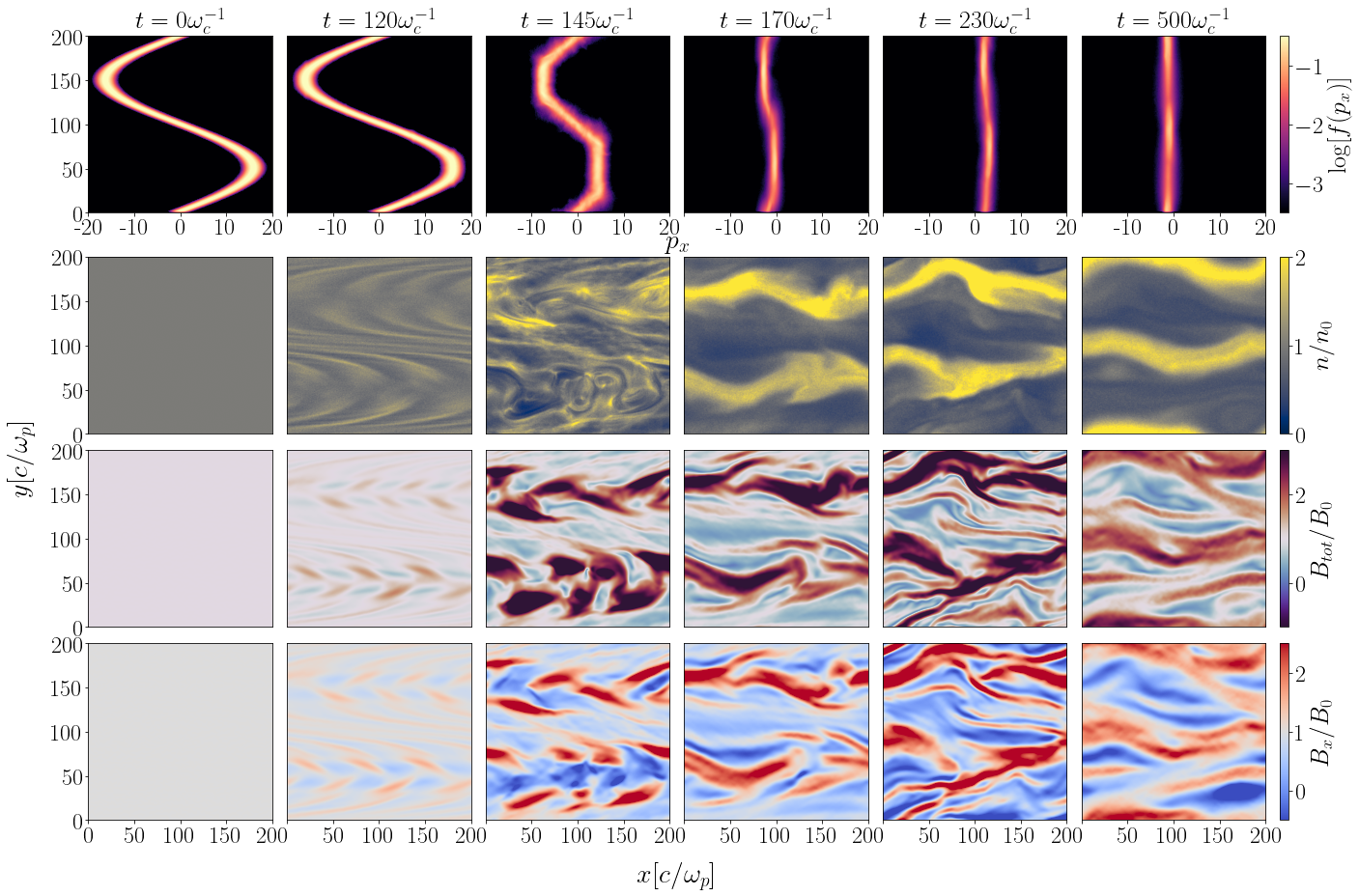}
\caption{From top to bottom: evolution of the Kolmogorov flow ($y-p_x$ phase space for thermal plasma), density $n$, total magnetic field $B_{tot}$, and $B_{x}$ over time for Run $\mathcal{B}$.
For the top row, the horizontal axis is the $p_{x}$ momentum of the thermal ions, in units of $m_iv_{A}$. 
For the other rows, the horizontal and vertical axes correspond to the box coordinates in units of $d_{i}$.}
\label{fig:evol}    
\end{figure*}

We first discuss our benchmark Run $\mathcal{B}$, which features a supersonic flow with $\MA\equiv U_0/v_A=20$, and CRs with number density $\ncr/\ngas=1\%$ and isotropic momentum  $\piso=200$;
hence, the CR energy density is about half of the kinetic energy density of the thermal plasma, which moves supersonically, and significantly larger than the background thermal one.

Besides a few trans/subsonic cases (runs $\mathcal{S}9$ and $\mathcal{S}10$), as in Paper I we focus on supersonic flows, which are more conducive to efficient particle acceleration and make it computationally easier to study nonthermal effects.

The supersonic/Alfv\'enic shear is inherently prone to disruption because the ordered magnetic field is dynamically unimportant ($M_A\gg1$).
The time evolution of the velocity phase space $y-p_x$, density, and magnetic fields for Run $\mathcal{B}$ are shown in Figure \ref{fig:evol}. 
As discussed extensively in Paper I, in supersonic flows  multiple physical processes take place simultaneously: on top of the Kelvin-Helmholtz instability, we have streaming instabilities and small-scale shocklets that disrupt the flow and grow into nonlinear fluctuations. 
The peaks in the velocity profile (at $y=50$ and 150) flatten out as particles start scattering (top row in Figure \ref{fig:evol}).
The region where the velocity shear is isotropized widens over time as turbulent eddies develop and transverse fluctuations in density and magnetic field develop and merge.
We observe an amplification of the magnetic field to $\sim 3-4B_0$ and density fluctuations of a factor of $\sim 2$, which correlate closely with the magnetic field strength, a signature of magnetosonic perturbations. 
After about 500 $\omega_{c}^{-1}$, the initial velocity structure is almost completely erased and the kinetic energy in the shear flow dissipated. 
We adopt the same characterization of the dissipation timescales that we introduced in Paper I. We consider the quantity
\begin{equation}
    \Delta(t)\equiv \frac{T_{xx}(t)}{T_{xx}(0)}; ~ T_{xx}(t) = \iint \rho(\mathbf{x}, t) v_x(\mathbf{x}, t)^2\, \mathrm{d}^2\mathbf{x}.
    \label{eqn:Delta}
\end{equation}
$T_{xx}$ is a proxy for the momentum flux, also proportional to the kinetic energy density, so $\Delta(t)$ represents the surviving fraction of shear momentum at time $t$. 

As in Paper I, we define three characteristic timescales $\tau_X$ based on  $\Delta(\tau_X)= X\%$:
(i) \textit{the onset time} $\tau_{90}$, when the shear coupling begins through kinetic instabilities;
(ii) \textit{the halving time} $\tau_{50}$, a general measure of shear dissipation; and
(iii) \textit{the viscous timescale} $\tau_\nu \equiv \tau_{20} - \tau_{90}$, which accounts for the bulk of the dissipation.
In the following we use $\tau_{90}$ when assessing the role of CRs in triggering the layer coupling and $\tau_{50}$ or $\tau_{\nu}$ to give the order of magnitude of the overall dissipation timescale.
\subsection{CR Viscosity}
\begin{figure}
\centering
\includegraphics[width=1\columnwidth]
{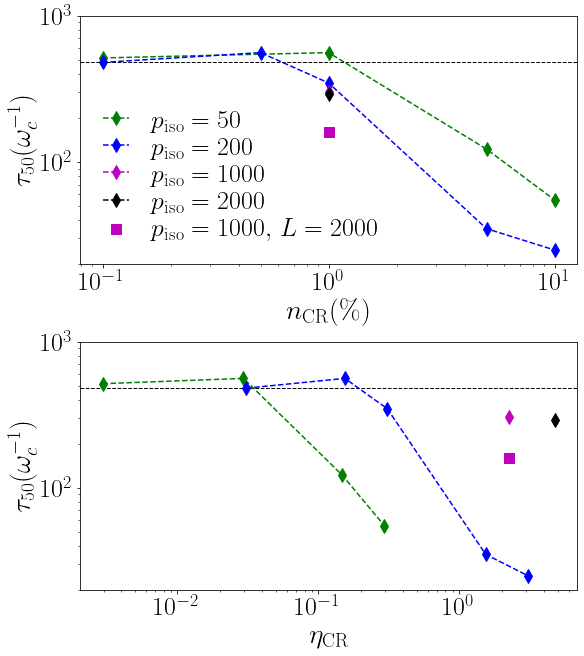}
\caption{Shear halving time $\tau_{50}$ as a function of CR number density (top) and kinetic energy density (bottom), for different $\piso$ as in the legend. 
The horizontal dashed lines correspond to a case without CRs. 
Above the thresholds of $\sim 1\%$ in $n_{\rm{CR}}$ and $\lesssim0.1$ in $\eta_{\rm{CR}}$, $\tau_{50}$ generally decreases when either $\ncr$ or $\piso$ increase (top panel), as well as when the CR energy density $\etacr$ increases (bottom).
The drop in $\tau_{50}$ saturates when the CR gyroradius becomes comparable to the shear scale.}
\label{fig:ncr_piso}
\end{figure}
We now focus on Runs $\mathcal{N}$, $\mathcal{P}$ and $\mathcal{E}$ in Table \ref{tab:param}, which test how $\tau_{50}$ depends on different CR number densities $\ncr$ and momenta $\piso$. 
We also introduce the ratio of energy density in CRs and thermal plasma
\begin{equation}
    \etacr \equiv \frac{\ncr(\gamma -1)  m_{i}c^2}{n_{\text{g}} m_{i}U_0^{2}}=\frac{\gamma -1}{M_{\rm{A}}^2}  \frac{\ncr}{n_{\rm{g}}}\frac{c^2}{v_{\text{A}}^{2}}
\end{equation}
where $\gamma$ is the Lorentz factor of CRs with momentum  $\piso$.
As mentioned above, such CR number and energy densities are constrained by numerical feasibility and should not  be strictly viewed as representative of astrophysical situations (e.g., Galactic CRs would have $\ncr/n_{\text{g}}\approx 10^{-9}$ and $\etacr\sim 1$ in the ISM).
Yet, they inform general scalings and are in the right ballpark to also represent populations of nonthermal particles accelerated by supersonic shears themselves, which typically have $\ncr/n_{\text{g}}\approx 0.01$ and $\etacr\sim 0.1$ (see Paper I).

Figure \ref{fig:ncr_piso} shows that both adding more CRs and making CRs more energetic lead to faster shear dissipation. 
As expected, CRs act as long-range messengers that connect relatively distant regions, promoting momentum exchange between different shear layers.
Quantitatively, Figure \ref{fig:ncr_piso} shows that there 
are thresholds ---around 1\% in $n_{\rm{CR}}$ and $\lesssim0.1$ in $\eta_{\rm{CR}}$--- above which the shear halving time significantly decreases with respect to the control run without CRs (dashed horizontal lines).
The trend is the same for $\piso=200$ and $\piso=50$ (blue and green dashed lines, respectively); 
for instance, adding 10\% of CRs with $\piso=200$ clears up the shear one order of magnitude faster compared to the case without CRs. 

At fixed CR number density $n_{\rm{CR}}=1\%$, increasing the CR $\piso$ leads to a slight decrease of $\tau_{50}$, but the effect saturates for $\piso\gtrsim 200$, which corresponds to CRs with gyroradius $r_g=\piso d_i\sim L$ in $B_0$, i.e.,  particles that cannot couple well with the shear. 
To confirm the importance of having $r_g<L$ for the CRs to exert viscosity, we performed Run $\mathcal{L}$ with ten times larger domain size and $\piso=1000$ (magenta square in Figure \ref{fig:ncr_piso}).
Though the shear flow with the same $U_0$ spans a larger domain (which should in principle slow down its dissipation), this configuration consistently extends the trend of $\tau_{50}$ being reduced for larger $\piso$.
Combining the effect of varying $\piso$ and $\ncr$ into a variation of $\etacr$ (right panel of Figure \ref{fig:ncr_piso}) we report a trend of a generally faster dissipation for larger CR energy densities, though not monotonic.
For example, 10\% of CRs of $\piso=50$ and 1\% of CRs of $\piso=200$ give roughly the same $\eta_{CR}$, but $\tau_{50}$ in the former is much shorter than the latter. 

In summary, while increasing the CR number density always induces faster shear dissipation, increasing the CR energy induces more viscosity only if their gyroradius remains smaller than or comparable with the typical shear lengthscale. 

\subsection{Particle Acceleration}
\begin{figure}
\centering
\includegraphics[width=1\columnwidth]{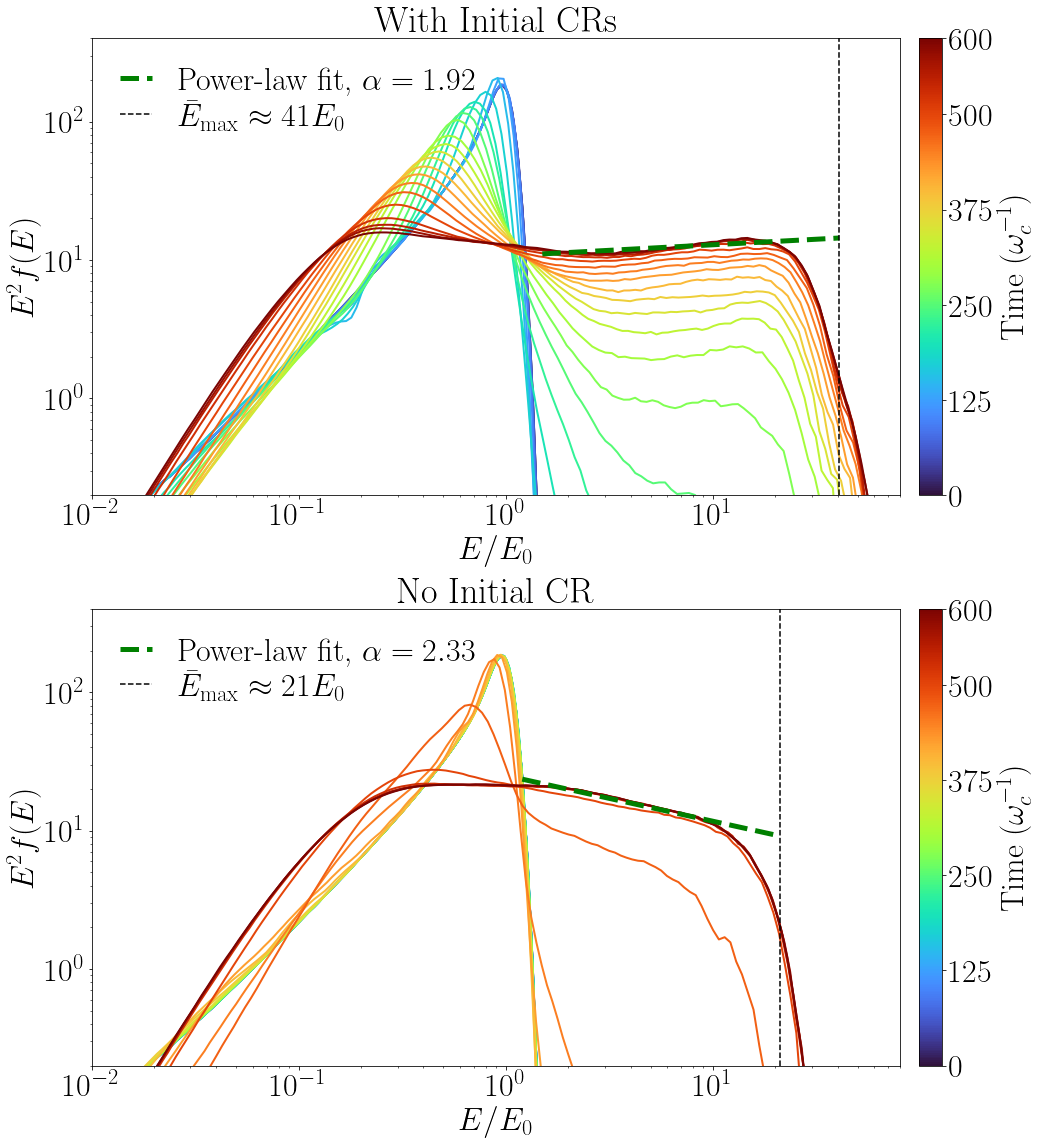}
\caption{
Evolution (color coded) of the energy spectra for the thermal ions in Run $\mathcal{B}$ and $\mathcal{N}5$ (top and bottom panel, respectively).
The legends provide the weighted maximum energies (black dashed lines) as $\Emax \sim 41 E_{0}$ for Run $\mathcal{B}$ and $\Emax \sim 21 E_{0}$ for Run $\mathcal{N}$5, consistent with the Hillas limit, and the slope of the power-law fits $\propto E^{-\alpha}$.
}
\label{fig:espec}
\end{figure}

Acceleration in a supersonic shear flow is generally an interplay of first/second-order Fermi processes of particles scattering between different shearing layers or local small-scale shocklets, and magnetic reconnection may also contribute.
Acceleration due to diffusion across shear layers has been studied before and is reviewed, e.g., in \citet{Rieger_2019};
also see references in Paper I.
In test-particle theory for a prescribed shear, stationary power-law spectra $n(p) \propto p^2 f(p) \propto p^{-(1+q)}$ may arise if the CR scattering length scales as a power-law in momentum as $\lambda_{\text{CR}(p)} \simeq c\tau \propto p^q$.
Still, we cannot expect this to hold in our setup because we also have the kinetic backreaction of energetic particles and the shear is dissipated on the acceleration timescale, so that  no stationary state can be achieved.

Figure \ref{fig:espec} shows that for Run $\mathcal{B}$ (top panel) the energy spectrum of background ions develops a nonthermal tail approximately $\propto E^{-2.2}$ (or $p^{-5.2}$, since particles are nonrelativistic), the extent of which grows in time.
Note that the ensuing spectra are appreciably steeper than the $\propto p^{-4}$ power-laws that one would expect for diffusive acceleration at strong shocks  \citep{bell78a, caprioli+14a}.

The bottom panel of Figure \ref{fig:espec} shows the evolution of the tail without pre-existing CRs (run $\mathcal{N}5$);
also in this case the ions develop a nonthermal tail, but acceleration starts later and achieves a smaller maximum energy than in Run $\mathcal{B}$. 
We conclude that CR seeds favor a significantly faster acceleration of the thermal ions, consistent with a faster onset of the shear dissipation (smaller $\tau_{90}$).
As a result, the maximum energy (vertical dashed line) at which thermal ions can be accelerated ends up being a factor of 4 times larger in the presence of CRs.

\begin{figure}
\centering
\includegraphics[width=\columnwidth]{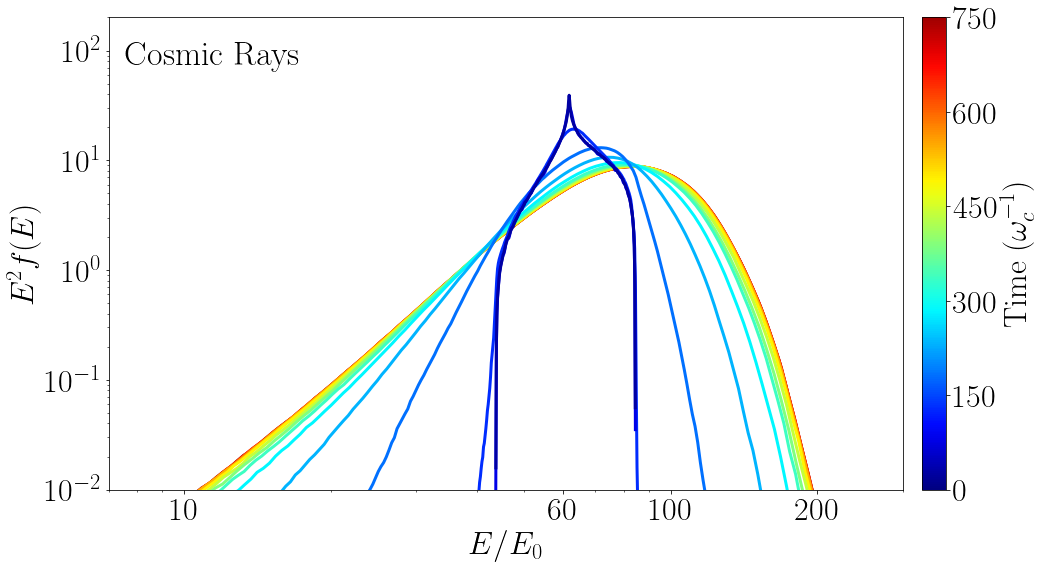}
\caption{Evolution of the CR energy spectrum in Run $\mathcal{B}$. The peak  moves up by a factor of $\sim2$ in energy, but the CRs show no significant sign of reacceleration for the finite box size and undriven shear considered here.}
\label{fig:crspecb}
\end{figure}

Figure \ref{fig:crspecb} shows the spectrum of CRs in Run $\mathcal{B}$;
initially monoenergetic, with time their distribution broadens and its peak moves to twice the initial energy; 
while some CRs gain up to a factor of $\sim10$, no power-law tail develops due to the finite box size (more on this in \S\ref{sec:Emax} when we discuss the Hillas limit). 

In general, power-law distributions arise from the balance of acceleration and escape (or loss) times \citep[][]{fermi54, bell78a}, though the latter  may be mimicked by time/space-dependence inducing acceleration bottlenecks. 
In our setup, the background shear continues to decay and is not homogeneous (Figure \ref{fig:evol}), so particles can be removed from the energization process by ending up in regions in which the shear has disappeared. 
In driven shears, with CRs steadily injected from the thermal pool, second-order Fermi acceleration behaves as expected as recently tested in MHD--PIC simulations by \citet{liu+26}.

\begin{figure}
\centering
\includegraphics[width=1\columnwidth]{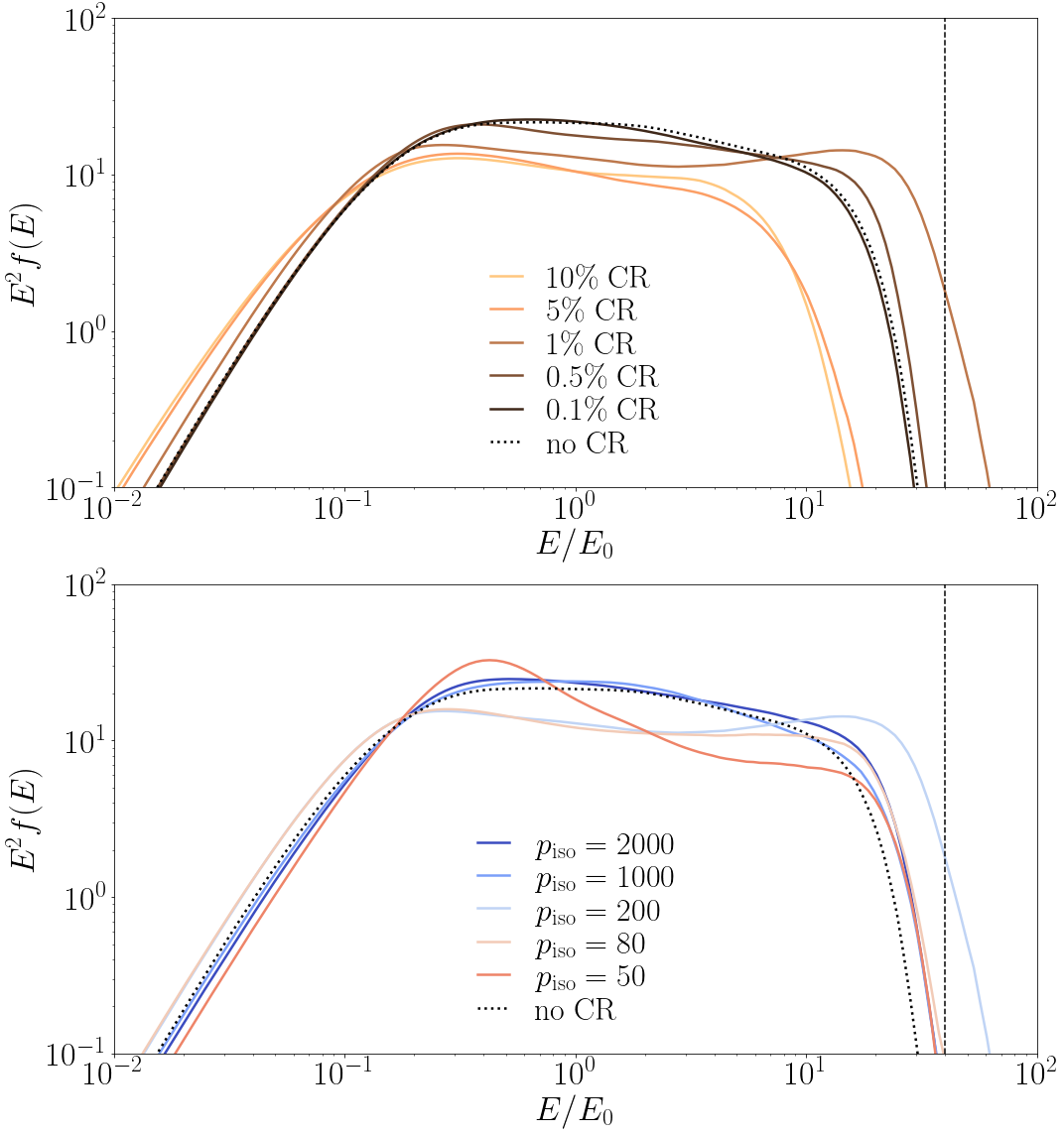}
\caption{Energy spectrum of thermal ions at $t=750\omega_{c}^{-1}$ (when the shear is reduced to $\Delta<20\%$), for $\piso=200$ and different $\ncr$ (top) and for $\ncr=1\%$ and different $\piso$ (bottom panel). The maximal particle energy approximately reaches the Hillas limit at $E_{\text{H}}\approx 40E_0$.}
\label{fig:specncr}
\end{figure}

To separate the effect of CR seeds, in Figure \ref{fig:specncr} we show the spectra of Run $\mathcal{B}$, $\mathcal{N}1-5$, and $\mathcal{P}1-4$ at $t=750\omega_{c}^{-1}$, i.e., when the shear has cleared ($\Delta<20 \%$) for all the cases. 
Quite surprisingly, not all runs with pre-existing CRs (e.g., the cases with $\ncr\gtrsim 5\%$) show more efficient particle acceleration than no-CR ones.
Similar to what we outlined for the shear reduction, adding CRs with gyroradius larger than the shear scale ($\piso\gtrsim 100$) has little effect on producing nonthermal ions (bottom panel of Figure \ref{fig:specncr}). 
We explain these trends by noticing that, when CR viscosity is prominent, there is a trade-off between rapid shear dissipation and sustained acceleration: when the shear is erased too quickly, the time available for ion energization also drops and nonthermal tails in the background ions are suppressed.
This is not expected to be the case for driven shears, though \citep[see][]{liu+26}.

\subsection{Maximum Energy}\label{sec:Emax}
\begin{figure}
\centering
\includegraphics[width=1\columnwidth]{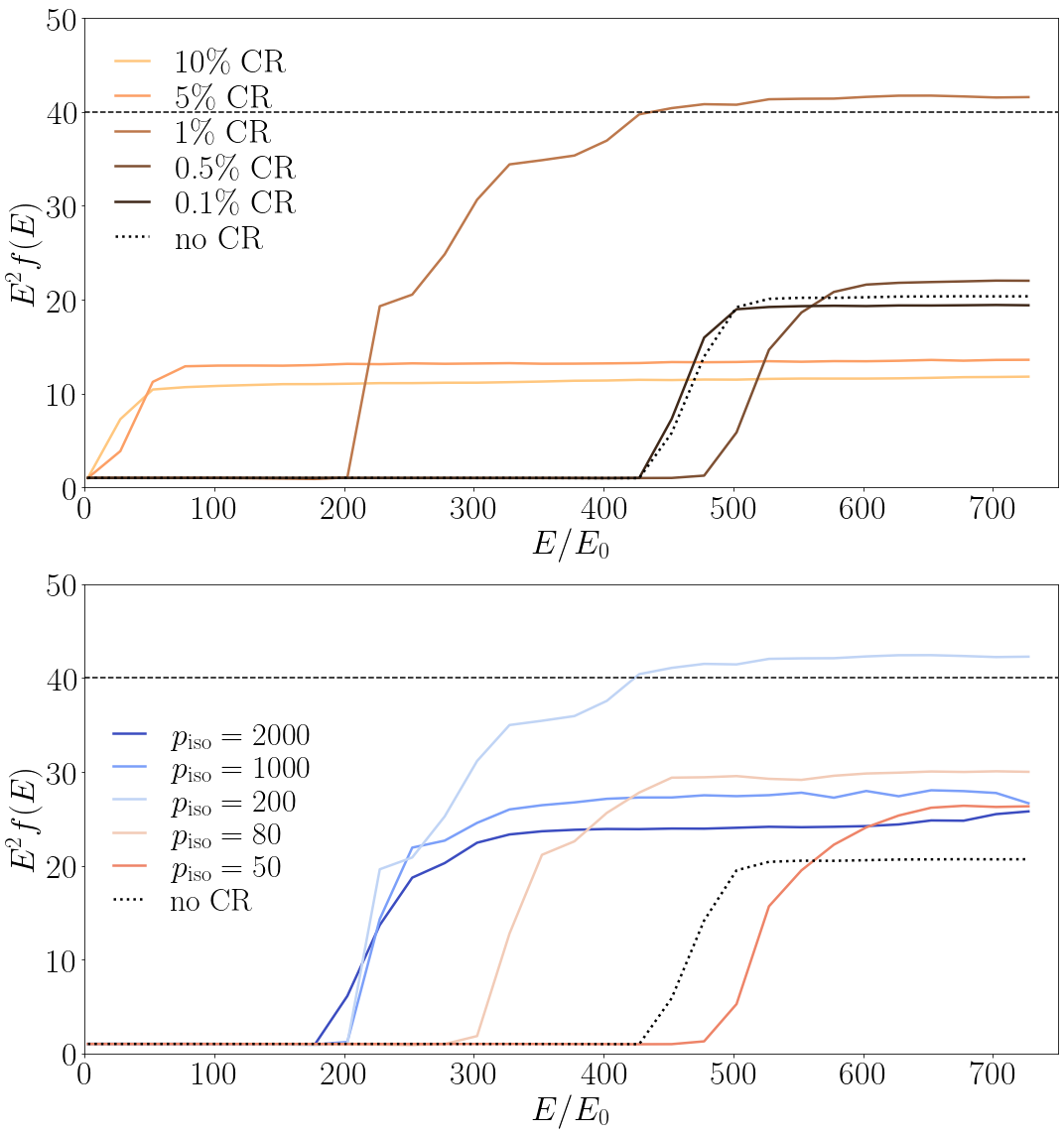}
\caption{Evolution of the weighted maximum particle energy $\Emax$ (Equation \ref{eq:Emax}) for different values of $\ncr$ (top)  and $\piso$ (bottom panel). The black dotted curve shows the case with no CRs, while the horizontal dashed lines correspond to the Hillas limit, $E_{\text{H}}\approx 40E_0$.}
\label{fig:emaxncr}
\end{figure}

We consider now the time evolution of the maximum  energy of the thermal ions.
Following \cite{bai+15}, we introduce a  weighted maximum energy defined as
\begin{equation}\label{eq:Emax}
\Emax\equiv\frac{\int E^{n+1} f(E) dE}{\int E^{n} f(E) dE}.
\end{equation}
For an energy distribution of the form $f(E)\propto E^{-m}\exp(-E/E_{\text{cut}})$, one obtains $\Emax\approx (n+1-m)E_{\text{cut}}$; we choose $n=6$, i.e., $\Emax\sim 5E_{\text{cut}}$.
It is useful to consider the limit energy determined by the Hillas criterion  \citep{hillas84}, $E_{\text{H}}\sim e U_0 B_0L/c$, i.e., the energy corresponding to the maximum potential drop of the motional electric field over the box size. 
For Run $\mathcal{B}$, we have $E_{\text{H}}\approx 40E_0$, consistent with the cut-off of the spectra of background ions (Figure \ref{fig:espec}) and the lack of extended tails in the CR spectra (Figure \ref{fig:crspecb}).

Figure \ref{fig:emaxncr} shows the evolution of $\Emax$ for runs with different $\ncr$ and $\piso$ (top and bottom panel, respectively).
The maximum energy scales with the parameters of the CR seeds similarly to the effective viscosity: adding CRs generally boosts $\Emax$, unless it induces a too fast suppression of the shear, in which case one gets diminishing returns. 
$\Emax$ also increases with $\piso$ until $\piso\gtrsim 100$, which corresponds to $\sim 42 E_0\sim E_{\rm H}$, the scale beyond which CRs have a gyroradius too large to couple with the shear, and then decreases.
All the $\Emax$ in Figure \ref{fig:emaxncr} are somehow below the benchmark case (and the Hillas limit), either because the CR are too few to have a role ($\ncr\lesssim 1\%$) or because they deplete the shear too quickly ($\ncr\gtrsim 1\%$). 
These parameters should not be taken as absolute values in assessing the role of CR seeds: they rather highlight the trade-off between the acceleration rate, which depends on the amount of turbulence and so increases with more CRs, and a finite acceleration time, which is limited by the survival of the shear.

\subsection{Free Kinetic Energy Partitioning}
\label{subsec:partition_cr}

\begin{figure}
\centering
\includegraphics[width=\columnwidth]{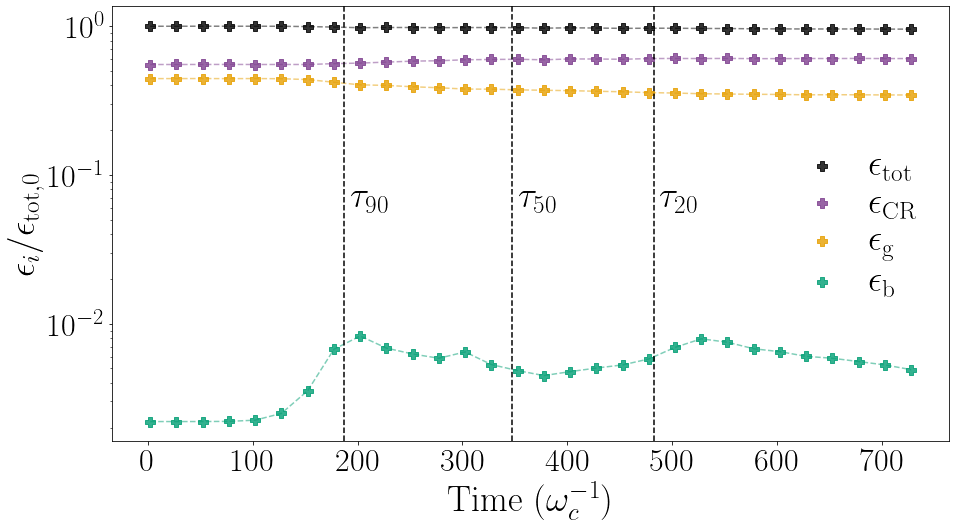}
\caption{Normalized energy densities in the thermal gas ions (both thermal and kinetic),  CR seeds, and magnetic fields. The total energy $\epsilon_{\text{tot}}$ is conserved within a few \%.}
\label{fig:efrac}
\end{figure}

In general, the initial free energy is partitioned among the thermal plasma, CRs, and magnetic fields, respectively labeled with $i=g, \rm{CR}$, and $B$.
Figure \ref{fig:efrac} shows how the fractions of the energy density in each species $\epsilon_i$ evolve over time for the benchmark Run $\mathcal{B}$.
Since the thermal gas distribution is a drifting Maxwellian, $\epsilon_g$ does not distinguish between proper thermal energy, kinetic energy, and nonthermal energy;
it is expected to decrease with time when the shear is dissipated and be converted into nonthermal components. 
We observe a slight increase in $\epsilon_{\rm CR}$ and a substantial increase in $\epsilon_B$, though the fraction of the energy in magnetic perturbations remains $\lesssim 1\%$.

\begin{figure}
\centering
\includegraphics[width=1\columnwidth]{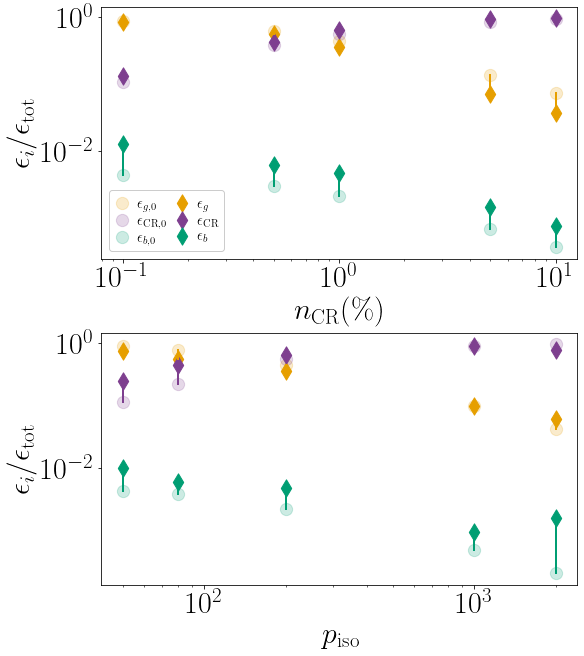}
\caption{Energy partitioning for different CR number densities and momentum. 
Diamonds and circles show the fractions of thermal ions, CRs, and magnetic field at final and initial times, respectively.
For both $\ncr$ and $\piso$, we have two cases initially dominated by CRs, one with comparable gas and CR energy, and two by the thermal plasma.}
\label{fig:efrac_np}
\end{figure}

Figure \ref{fig:efrac_np} illustrates how the energy partitioning (at $t=750 \omega_{c}^{-1}$, when most of the shear is dissipated) changes with varying $\ncr$ and $\piso$ (runs $\mathcal{B}$, $\mathcal{N} 1-5$, and $\mathcal{P} 1-4$).
In most cases the thermal plasma loses energy, i.e., the free kinetic energy does not end up in heating or acceleration of the thermal component itself, but it is rather converted into CRs and magnetic fields;
this purely kinetic effect is represented by the displacement of the markers with respect to the corresponding initial value (faint markers).
Our survey shows that, as $\piso$ increases, the energy loss from the thermal gas is slightly reduced. Notably, in Run $\mathcal{P}$1, where the initial $\ecr\gtrsim 95\%$ of the total, even if the initial kinetic energy of the gas is dissipated, $\epsilon_{g}$ grows and the thermal ions gain energy at the expense of the CRs.
However, note that runs with high $\ncr$ ($\mathcal{N}$1 and $\mathcal{N}$2) have initial $\ecr$ close to 90\% as well, but in these cases energy still flows from the thermal gas to the CRs. 
Such a difference cannot be ascribed to the role of the CR viscosity, which is larger for both higher $\ncr$ and $\piso$, but rather suggests that the partitioning between thermal plasma and CR energies is not a mere thermodynamical equilibration, but a kinetic process controlled by the ability of CRs with adequate gyroradii to extract energy from the decaying shear.

\section{Scaling with the Shear Mach Number}\label{sec:sigmas}
As shown in Paper I, steeper velocity gradients in a shear flow generally lead to faster shear reduction. 
Here, we test different shear velocities, with Alfv\'en Mach numbers $M_A\equiv U_0/v_A$ between 0.5 and 40 (runs $\mathcal{S}1$--$\mathcal{S}10$).
All the other CR and box parameters are kept fixed as in the benchmark.

\subsection{Dissipation Timescale}
\begin{figure}
\centering
\includegraphics[width=\columnwidth]{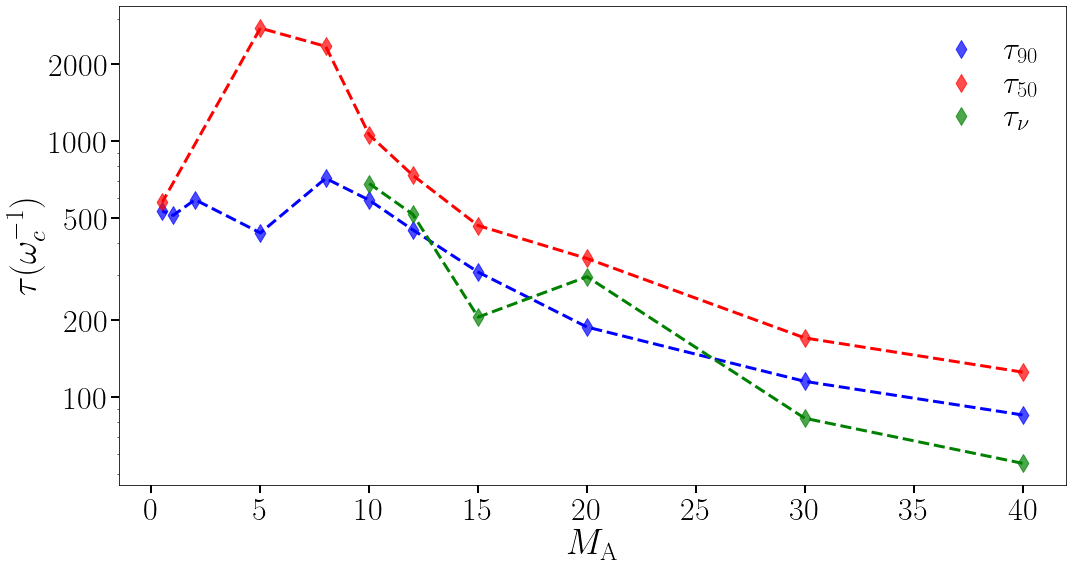}
\caption{The shear reducing time ($\tau_{90}$, $\tau_{50}$ and $\tau_{\nu}$) for Run $\mathcal{S}1-6$ with different initial Alfv\'enic Mach numbers. The free kinetic energy is reduced faster in faster shear flows.}
\label{fig:tau50_Ma}
\end{figure}

Figure \ref{fig:tau50_Ma} shows the shear reducing timescales for different shearing velocities (Run $\mathcal{S}1-\mathcal{S}10$). 
Below $\MA=10$, the onset time $\tau_{90}$ is roughly constant; 
all three timescales decrease as the initial velocity gradient gets larger at greater initial $\MA\gg 1$. 
A shorter $\tau_{50}$ indicates a larger viscosity that includes both the effect of accelerated particles and the pre-existing CRs. 
$\Delta$ hardly drops below 20\% of its initial value in $\mathcal{S}6-\mathcal{S}10$, i.e., for $\MA\lesssim 5$, hence the missing points in the $\tau_\nu$ curve;
for $\MA=2$ and $\MA=1$ saturation occurs at even higher values of $\Delta \gtrsim 0.5$.
This is the effect of the magnetic field on the KH instability: unlike in Paper I, where $\mathbf{B}_0$ was perpendicular to the flow, the relatively strong component of $\mathbf{B}$ along the shear inhibits further dissipation.

\subsection{Acceleration Efficiency}
Let us consider now how the acceleration of thermal background particles depends on $\MA$.
As in Paper I, we define a reference energy 
\begin{equation}\label{eq:E0}
    E_{0}\equiv \frac{1}{2}m_{i}U_0^2+\frac{3}{2}m_{i}v_{\rm th,i}^2,
\end{equation}
which includes both the bulk shear kinetic energy and the thermal energy, and label \textit{nonthermal} the particles that are accelerated beyond $2E_0$.

\begin{figure}
\centering
\includegraphics[width=\columnwidth]{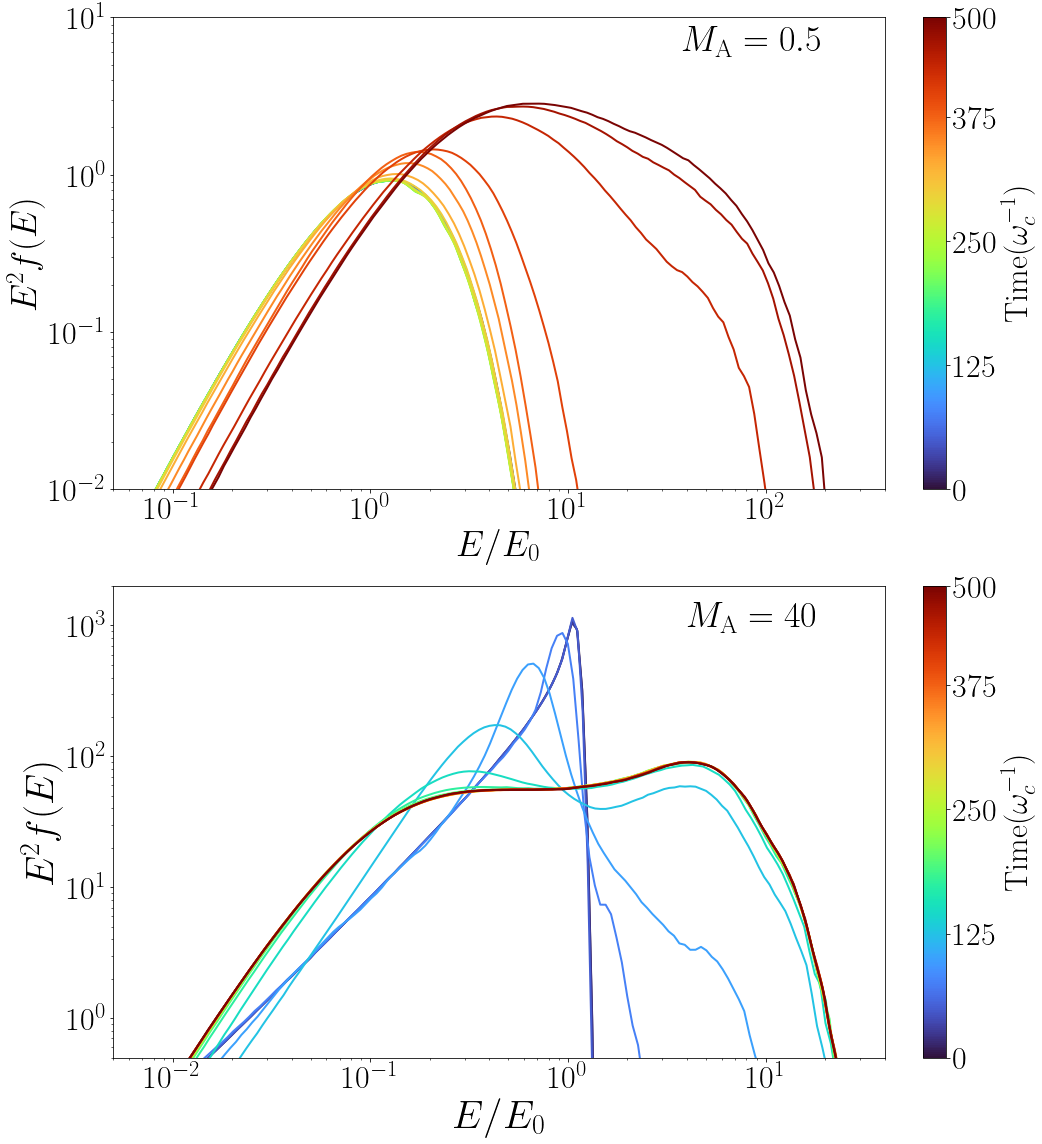}
\caption{Evolution of energy spectra of the background plasma for Mach numbers $M_{\rm{A}} = 0.5$, and $M_{\rm{A}} = 40$ (runs $\mathcal{S}10$ and $\mathcal{S}1$, top and bottom, respectively) color coded by time. 
In the subsonic case, thermal ions are heated and develop a steep nonthermal tail, whereas in the very supersonic case, they are accelerated much more efficiently with a hard nonthermal spectrum that carries most of the energy.}
\label{fig:espec_evol}
\end{figure}

Figure \ref{fig:espec_evol} shows the evolution of the energy spectra of the thermal background ions for runs $\mathcal{S}1$ (supersonic), and $\mathcal{S}10$ (subsonic). 
At low Mach numbers ($M_{\rm{A}}=0.5$), particles are heated up and the energy peak shifts to higher energies, reaching $\sim10E_0$ over time; 
a steep tail at energies as high as $\sim100E_0$ also develops.
At larger $M_{\rm{A}}=40$, instead, particles rapidly gain energy and develop a hard nonthermal tail, where most of the energy is stored, indicative of efficient acceleration. 

\begin{figure}
\centering
\includegraphics[width=\columnwidth]{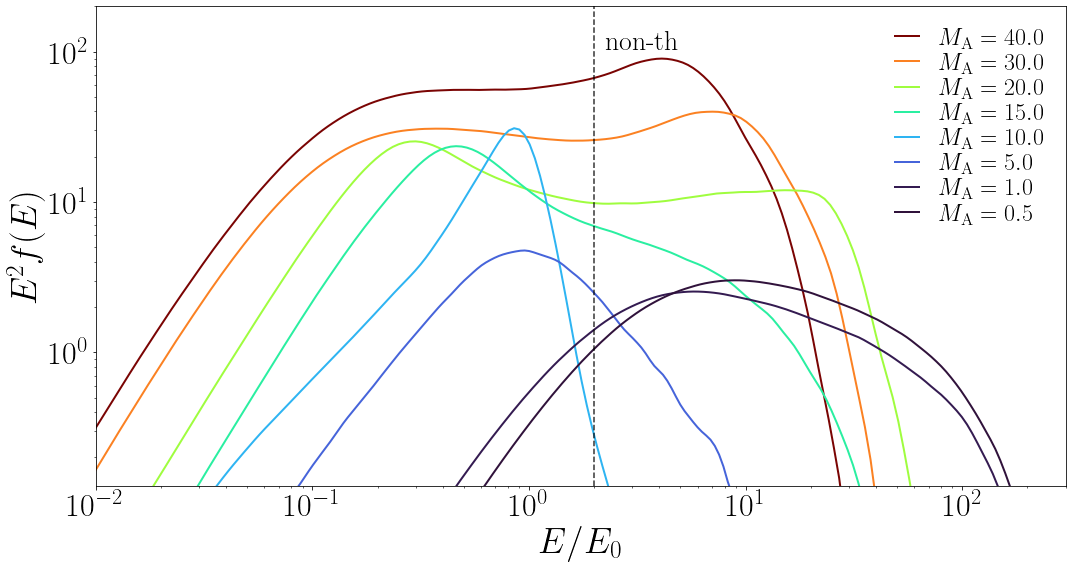}
\includegraphics[width=\columnwidth]{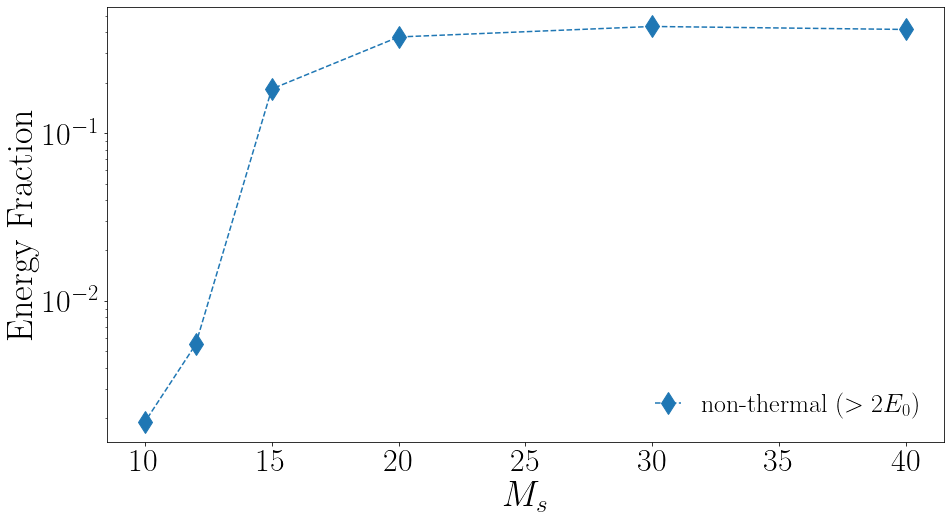}
\caption{Top panel: Energy spectra of the background ions at t=$625\omega_{c}^{-1}$, with energy normalized as in Equation \ref{eq:E0}, as a function of $\MA$. 
Ions with $E \ge 2E_{0}$ are labeled as \textit{nonthermal}.
Bottom panel: Energy fraction in nonthermal ions at $\tau_{20}$ for Run $\mathcal{S}1-5$.
The acceleration efficiency increases with $M_{\rm{A}}$ and saturates at $\sim 40\%$ for $M_A\gtrsim 20$.}
\label{fig:cos_spec}
\end{figure}

Figure \ref{fig:cos_spec} presents the energy spectra of the thermal gas ions at $t=625 \omega_{c}^{-1}$ for simulations with different $\MA$ (top panel), along with the fraction of energy density in nonthermal ions for highly supersonic runs ($\mathcal{S}1-5$) in the bottom panel.
Overall, the evolution of the energy spectrum across different Mach numbers reflects a transition from heating (i.e., the shift of the bulk of the particles to higher energies) to the development of nonthermal distributions at high Mach numbers.
Both low- and high-$\MA$ cases show nonthermal tail extending about one order of magnitude, but spectra are much flatter for more supersonic cases. 
The energy fraction in such nonthermal particles increases from less than one percent to about 40\% of the total for $\MA\gtrsim 10$ (bottom panel of Figure \ref{fig:cos_spec}).
To better understand these trends, it is necessary to consider the actual source of free energy in the different cases, which is the topic of the next section.

\subsection{Energy Partitioning}
\label{subsec:efrac}
\begin{figure}
\centering
\includegraphics[width=\columnwidth]{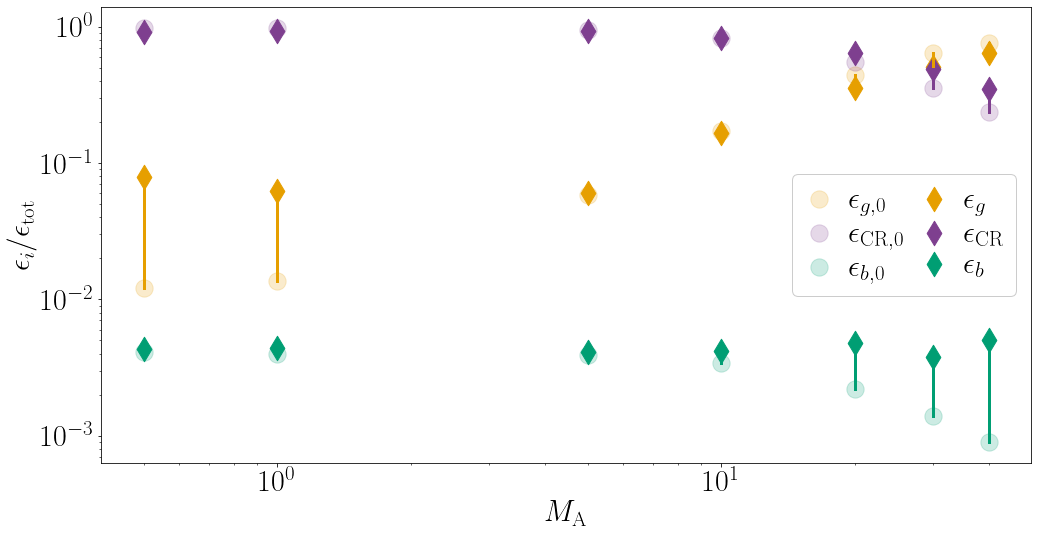}
\caption{Energy density $\epsilon_{i}$ in species $i$ at t=750 $\omega_{c}^{-1}$ for different initial $\MA$. The red, blue, and green diamonds show the energy density fraction of thermal gas, CRs, and magnetic field respectively, while the fainter dots of the same color show them at the initial time step.}
\label{fig:eps_Ma}
\end{figure}

We discuss now the free energy partitioning into thermal plasma, CRs, and magnetic fields (as in \S\ref{subsec:partition_cr}), as a function of the shear Mach number.
For our benchmark parameters, thermal plasma and CR seeds are in equipartition at $\MA=20$; 
hence, for $\MA\lesssim 20 $ the energy budget is dominated by CRs, while for $\MA\gtrsim 20$ by the thermal plasma.
Interestingly, $\epsilon_B\sim 5\times 10^{-3}$, basically independent of $\MA$;
though for $\MA\lesssim 20 $ this corresponds to the initial value of $B_0$, for larger $\MA$ it is the result of field amplification.
$\epsilon_g$ generally increases with $\MA$, but it evolves in a very different way depending on the regime.
For sub/trans-Alfv\'enic cases, we observe an increase of about one order of magnitude in $\epsilon_g$, at the expense of the CRs, which are the only species with larger energy density;
in these cases the shear is not the energy donor, but rather the agent triggering the fluctuations that couple  thermal particles with CRs, favoring energy redistribution. 
For $\MA\gtrsim 20$ the trend discussed in \S\ref{subsec:partition_cr} is confirmed: both thermal particles and CRs can gain energy at the expense of the shear. 
We note that, while such an energy redistribution may apply also to driven shears, the saturation values may differ substantially, and will be the subject of a future study. 

\section{Conclusion}\label{sec:conclusion}
In Paper I \citep{liang+26} we performed hybrid simulations of decaying  subsonic and supersonic shears to investigate how the free kinetic energy is dissipated into heat, magnetic fields, and accelerated particles.
In this second work, we considered the effects of a pre-existing population of energetic particles (CR seeds) on the effective shear viscosity and studied how their presence affects the energy partitioning and the acceleration of background particles. 
Our main conclusions can be summarized as follows:

\begin{itemize}
\item  Pre-existing CRs generally act as long-range messengers able to connect distant shear regions \citep{earl+88}. 
CRs with gyroradii smaller than the shear extent couple well with the self-generated magnetic fluctuations and foster momentum exchange, effectively creating a \textit{CR viscosity} that speeds up the decay of the velocity structure.
 
\item CRs can exert an effective viscosity even when they are energetically underdominant; with larger CR number density or momentum, the shear is reduced faster (Figure \ref{fig:ncr_piso}). 
The trend is monotonic with the energy density in CRs, as long as the CR gyroradius remains smaller than the shear scale-height. 
When this condition is violated, CRs do not couple effectively with the thermal plasma and their presence does not affect either the shear evolution or the acceleration of thermal ions.
These results are consistent with the findings of Paper I, where the nonthermal particles were not seeded but spontaneously generated by the shear itself.

\item 
Pre-existing CRs also affect the acceleration of background particles and the partition of the free kinetic energy, which can be channeled into thermal ions,  CR seeds, and magnetic fluctuations.
Energy transfer between the thermal gas and the CRs can go both ways, depending on their initial predominance (Figure \ref{fig:eps_Ma}), though this does not change the importance of CR viscosity for the shear evolution.
These kinetic effects are important at the leading order and cannot be captured by fluid or MHD approaches. 

\item Thermal ions are both heated and accelerated during the shear dissipation, with the process happening much faster and developing flatter power-law tails in the presence of CRs (Figure \ref{fig:espec}). 
The maximum energy achievable by such particles depends on the amount of CRs in a nontrivial way: on one hand, CRs speed up acceleration, but on the other hand their viscosity depletes the shear more rapidly. 
The competition between a faster acceleration rate and a reduced shear lifetime sometimes does not allow CRs to achieve the Hillas limit (Figures \ref{fig:emaxncr} and \ref{fig:specncr}).

\item Magnetic turbulence is invariably produced by the decay of the shear, typically encompassing about 1\% of the initial free energy for our undriven simulations (Figure \ref{fig:efrac_np}). 
Though well below equipartition with CRs and background plasma, these self-generated fluctuations are key to dynamically couple all the species. 

\item Some of the results above depend strongly on the shear not being driven: in the presence of continuous energy injection we expect the power-law tails of the background ions to extend up to the Hillas limit and the magnetic turbulence to grow close to equipartition.  
\end{itemize}

This series aims to characterize the role of kinetic effects in shearing, collisionless, nonrelativistic plasmas, particularly in supersonic flows and in the presence of CR seeds. With the extrapolations needed to connect kinetic simulations to astrophysical scales, our results suggest that viscosity induced by nonthermal particles---whether accelerated or pre-existing---is an intrinsic property of collisionless astrophysical plasmas. Future work will explore in more detail how CR viscosity mediates momentum transfer, amplifies magnetic fields, and accelerates CRs in interstellar and intracluster turbulence, shearing layers (e.g., jets or stellar winds), and accretion disks.

\begin{acknowledgments}
We thank Thomas Berlok, Ellen Zweibel, Peng Oh, Mateusz Ruszkovski, Anatoly Spitkovksy, Mingxuan Liu, Xiaochen Sun, and Tsun Hin Navin Tsung for helpful discussions on CR viscosity in shearing layers. We also thank the University of Chicago Research Computing Center for computational resources. This work was supported in part by NASA grant 80NSSC18K1726, NSF grants AST-2510951 and AST-2308021 to D.C., NSF grant PHY-2309135 to the Kavli Institute for Theoretical Physics, and NSF grant AST-240752 to N.L.
\end{acknowledgments}


\bibliographystyle{aasjournalv7}
\end{document}